\documentclass[review,number,sort&compress]{elsarticle}
\usepackage{tikz}
\usepackage{float}
\usepackage{graphicx}
\usepackage{gensymb}
\usepackage{amssymb}
\usepackage{color,soul}
\usepackage{url}
\usepackage{ulem}
\usepackage{textcomp}
\usepackage[figuresright]{rotating}
\usepackage{subcaption}
\usepackage{lineno}
\usepackage{hyperref}
\usepackage{cleveref}
\usepackage{libertine}
\usepackage{siunitx}
\usepackage[a4paper, total={5.0in, 8in}]{geometry}
\usepackage[T1]{fontenc}
\usepackage{tablefootnote}

\definecolor{amethystbg}{rgb}{0.6, 0.4, 0.8}
\definecolor{coolgreybg}{rgb}{0.55, 0.57, 0.67}
\definecolor{babypinkbg}{rgb}{0.96, 0.76, 0.76}
\definecolor{cadmiumgreenbg}{rgb}{0.0, 0.42, 0.24}
\definecolor{bluebg}{rgb}{.63,.79,.95}
\definecolor{orangebg}{rgb}{1,0.5,0}
\colorlet{lightbluebg}{bluebg!40}
\colorlet{lightorangebg}{orangebg!40}
\colorlet{lightcadmiumgreenbg}{cadmiumgreenbg!40}
\colorlet{lightbabypinkbg}{babypinkbg!40}
\colorlet{lightcoolgreybg}{coolgreybg!40}
\colorlet{lightamethystbg}{amethystbg!40}

\DeclareGraphicsExtensions{.eps,.pdf,.png,.jpg,}

\journal{Nuclear Instruments and Methods A}

\begin{document}

\newcommand{\etal}{{\it et~al.}}
\newcommand{\geant} {{{G}\texttt{\scriptsize{EANT}}4}}
\newcommand{\srim} {\texttt{SRIM}}
\newcommand{\python} {\texttt{Python}}
\newcommand{\pandas} {\texttt{pandas}}
\newcommand{\SciPy} {\texttt{SciPy}}
\newcommand{\ROOT} {\texttt{ROOT}}

\DeclareRobustCommand{\hlb}[1]{{\sethlcolor{lightbluebg}\hl{#1}}}
\DeclareRobustCommand{\hlo}[1]{{\sethlcolor{lightorangebg}\hl{#1}}}
\DeclareRobustCommand{\hlg}[1]{{\sethlcolor{lightcadmiumgreenbg}\hl{#1}}}
\DeclareRobustCommand{\hlp}[1]{{\sethlcolor{lightbabypinkbg}\hl{#1}}}
\DeclareRobustCommand{\hlgr}[1]{{\sethlcolor{lightcoolgreybg}\hl{#1}}}
\DeclareRobustCommand{\hla}[1]{{\sethlcolor{lightamethystbg}\hl{#1}}}

\begin{frontmatter}


\title{Light-yield response of liquid scintillators using $2-6$ MeV tagged neutrons\tnoteref{label1}}

\author[lund]{N.~Mauritzson}
\author[lund,ess]{K.G.~Fissum\corref{cor1}}
\author[glasgow]{J.R.M.~Annand}
\author[lund]{H.~Perrey}
\author[ess,glasgow]{R.~Al~Jebali}
\author[ess,glasgow]{A.~Backis}
\author[ess,glasgow,trento]{R.~Hall-Wilton}
\author[ess]{K.~Kanaki}
\author[lund,ess]{V.~Maulerova-Subert\fnref{fn2}}
\author[lund]{F.~Messi\fnref{fn3}}
	\author[lund,uppsala]{R.J.W.~Frost}
\author[lund]{E.~Rofors\fnref{fn5}}
\author[thermofisher]{J.~Scherzinger}
	
\address[lund]{Division of Nuclear Physics, Lund University, 221 00 Lund, Sweden}
\address[ess]{Detector Group, European Spallation Source ERIC, 221 00 Lund, Sweden}
\address[glasgow]{School of Physics and Astronomy, University of Glasgow, Glasgow G12 8SU, United Kingdom} 
\address[trento]{Sensors \& Devices Centre, Fondazione Bruno Kessler, via Sommarive 18, 38123 Trento, Italy}
\address[uppsala]{Division of Applied Nuclear Pnysics, Uppsala University, 751 20 Uppdala, Sweden}
\address[thermofisher]{Thermo Fisher Scientific,  Industrial Park Frankfurt Hoechst, 65926 Frankfurt am Main, Germany}
 
\tnotetext[label1]{The data set doi:10.5281/zenodo.10053068 is available for download from \href{https://zenodo.org/record/10053068}{https://zenodo.org/record/10053068.}}

\cortext[cor1]{Corresponding author. E-mail: kevin.fissum@nuclear.lu.se}
 
    \fntext[fn2]{Present address: CERN, European Organization for Nuclear Research, 1211 Geneva, Switzerland and Hamburg University, 20148 Hamburg, Germany}
    
    \fntext[fn3]{Present address: Svensk Kärnbränslehantering AB, Evenemangsgatan 13, Box 3091, 169 03 Solna, Sweden}
    
    
    \fntext[fn4]{Present address: Lawrence Berkeley National Laboratory, 1 Cyclotron Rd, Berkeley, California 94720, United States}
    
\begin{abstract}
Knowledge of the neutron light-yield response is crucial to the understanding of scintillator-based neutron detectors. 
In this work, neutrons from 2--6\,MeV have been used to study the scintillation light-yield response of the liquid scintillators NE~213A, EJ~305, EJ~331 and EJ~321P using event-by-event waveform digitization. 
Energy calibration was performed using a \geant~model to locate the edge positions of the Compton distributions produced by gamma-ray sources. 
The simulated light yield for neutrons from a PuBe source was compared to measured recoil proton distributions, where neutron energy was selected by time-of-flight. 
This resulted in an energy-dependent Birks parametrization to characterize the non-linear response to the lower energy neutrons. 
The NE~213A and EJ~305 results agree very well with existing data and are reproduced nicely by the simulation. 
New results for EJ~331 and EJ~321P, where the simulation also reproduces the data well, are presented.
 
\end{abstract}


\end{frontmatter}
 
\newpage

\section{Introduction}
\label{section:Introduction}
The detection of fast neutrons in fields of gamma-rays is often accomplished using organic liquid scintillators. 
Knowledge of the light-yield response of these organics is important for the understanding of the neutron and gamma-ray detection mechanism.
The organic liquid scintillator NE~213~\cite{ne213} and its more recent derivative NE~213A~\cite{ANNAND1997} have been used widely~\cite{BATCHELOR196170}. 
The performance of these organics is often employed as a benchmark in the development of fast-neutron detector materials and systems~\cite{BAYAT2012217,IWANOWSKA201334,PAWELCZAK201321,JEBALI2015102}. 
Newer liquid scintillators include the high scintillation-light yield EJ~305~\cite{ej305} and EJ~309~\cite{ej309}, EJ~331~\cite{ej331} (which includes a thermal-neutron sensitive gadolinium additive), and EJ~321P~\cite{ej321p} (a mineral-oil based scintillator with a 2:1 hydrogen:carbon ratio). 
Recently, a \geant~model~\cite{AGOSTINELLI2003250, 1610988} was developed to facilitate the gamma-ray energy calibration \cite{MAURITZSON2022165962} of these types of detectors. 
Here, this \geant~model was extended to include the neutron scintillation-light yield with an energy-dependent Birks parameter. 
A polychromatic neutron source and the time-of-flight (TOF) technique were employed to measure the scintillator responses as a function of incident neutron energy.
The simulated neutron scintillation yield corresponding to the maximum neutron-energy deposition was compared to the measured scintillation yield at the edge of the recoil-proton distribution. 
This edge corresponds to all of the kinetic energy of the incident neutron being transferred to a scintillator hydrogen atom in a single collision. 
In this paper, a detailed study of the light yield of the NE~213A, EJ~305, EJ~331 and EJ~321P scintillators is presented. 
Results for NE~213 and EJ~305 are compared with previous studies and first results are presented for EJ~331 and EJ~321P.
The excellent agreement between the simulated neutron scintillation-light yield and the data highlights the detailed understanding of the underlying scintillation mechanisms and light-collection processes.

\section{Apparatus}
\label{section:Apparatus}

\subsection{PuBe-based neutron and gamma-ray source}
\label{subsection:ApparatusPuBeSource}

A $^{238}$Pu/$^{9}$Be (PuBe) source provided the fast neutrons. $^{238}$Pu decays via $\alpha$-particle emission to $^{234}$U producing $\alpha$ particles of the energy $\sim$$5.5$\,MeV~\cite{nudat}. A cascade of low-energy gamma-rays is emitted from the subsequent de-excitation of $^{234}$U to the ground state. $\alpha$-particles which interact with $^{9}$Be via the $\alpha$~$+$~$^{9}$Be~$\rightarrow$~$^{12}$C~$+$~n reaction have a maximum kinetic energy of $\sim$11\,MeV when the recoiling $^{12}$C is left in the ground state.
When the recoiling $^{12}$C is left in the first-excited state, a 4.44\,MeV gamma-ray is emitted from the subsequent de-excitation.
This occurs $\sim$50\% of the time. 
Thus, the radiation associated with PuBe includes fast neutrons with energies up to $\sim$11\,MeV, low-energy cascade gamma-rays and energetic 4.44\,MeV gamma-rays. 
Energy conservation restricts the maximum energy of neutrons emitted in coincidence with a $4.44$\,MeV gamma-ray to $\sim$$6$\,MeV. 
The neutrons are ``tagged'' if both particles are detected, as the coincident 4.44\,MeV gamma-ray provides a reference for a TOF measurement. 
The PuBe source emitted $\sim$2.9~$\times$~10$^6$ neutrons per second~\cite{radiochemicalcentre} nearly isotropically, see Ref.~\cite{SCHERZINGER201798}.

\subsection{Detectors}
\label{subsection:ApparatusDetectors}

\subsubsection{Gamma-ray trigger detectors}
\label{subsubsection:ApparatusDetectorsGammaRayTriggerDetector}
In the MeV energy range, Yttrium Aluminum Perovskit:Cerium (Ce$^{+}$ doped YAlO$_{3}$, YAP:Ce) inorganic crystals~\cite{MOSZYNSKI1998157} have good gamma-ray detection efficiency and low efficiency for neutrons. 
Four YAP:Ce detectors from Scionix~\cite{scionix} were used to detect both the low-energy cascade and 4.44\,MeV gamma-rays. 
The cylindrical crystals were 1\,in. $\times$ 1\,in. (diameter $\times$ height) and were attached to a 1\,in. Hamamatsu Type R1924 photomultiplier tube (PMT)~\cite{hamamatsu}. 
Gamma-rays from a $^{22}$Na source ($E_{\gamma}$~=~1.28\,MeV) were used to set the gains of the YAP:Ce detectors at an operating voltage of about $-$750\,V.

\subsubsection{Fast-neutron/gamma-ray detectors}
\label{subsubsection:ApparatusDetectorsFastNeutronGammaRayDetector}
The liquid scintillators were contained in identical cylindrical aluminum cells (94\,mm in diameter $\times$ 62\,mm deep, $\sim$430\,cm$^{3}$ detection volume, wall thickness 3\,mm). 
A TiO$_2$-based reflector (EJ~520~\cite{ej520}) coated the inside of each cell. 
Optical windows consisted of 5\,mm thick borosilicate glass disks~\cite{borosilicate} glued to each cell using Araldite 2000$+$~\cite{araldite}.
The cells were filled through ports which were then sealed with Viton O-rings~\cite{viton} compressed with aluminum screws. 
The cells were dry fitted (without optical coupling medium) to a cylindrical PMMA UVT~\cite{pmma} lightguide (72.5\,mm in diameter $\times$ 57\,mm long). TiO$_2$-based reflector (EJ~510~\cite{ej510}) was used to coat the external curved surfaces of the light guide and each assembly was dry fitted to a 3\,in. diameter Electron Tubes type 9821KB PMT ~\cite{et_9821kb}. 
A set of springs was used to hold the cell, lightguide and PMT face in contact and a mu-metal magnetic shield was fitted around the PMT. 
The PMTs were operated at about $-2$\,kV, the voltage employed in previous investigations~\cite{SCHERZINGER201574, JEBALI2015102, JuliusScherzinger2016, SCHERZINGER201798}. 
The signal amplitudes were adjusted using variable attenuators (CAEN type N858~\cite{caen_n858}). 
Typical 1\,MeV$_{ee}$ signals had amplitudes of about $-$700\,mV, risetimes of $\sim$5\,ns and falltimes of $\sim$60\,ns. 

Four different liquid scintillators were employed (Table~\ref{table:ScintillatorProperties}): 

\setcounter{footnote}{0} 

\begin{itemize}
    \item NE~213A, a pseudocumene-based variant of the organic NE~213 developed specifically for neutron/gamma-ray discrimination.
    
    \item EJ~305, a pseudocumene-based organic similar to NE~224~\cite{MADEY1978445} and BC~505~\cite{bc505} with a high scintillation-light yield.
    
    \item EJ~331, a pseudocumene-based organic doped with gadolinium ($1.5$\% by weight)\footnote{The base scintillator is taken to have properties similar to EJ~309 throughout the remainder of this paper.}.
    
    \item EJ~321P, a mineral-oil based scintillator with a hydrogen-to-carbon ratio larger than 2.
\end{itemize}

\begin{table}[H]
    \centering
    \footnotesize
    \caption{Selected scintillator properties.}
    \begin{tabular}{r r r r r}\hline
        Scintillator                & NE~213A  & EJ~305 & EJ~331\tablefootnote{These properties correspond to the datasheet for EJ~331 ($0.5$ Gd\,\%w/w).}  & EJ~321P  \\ \hline 
        Density [g/cm$^3$]  & 0.87& 0.89 &0.89 & 0.85\\
        Light Yield (\% Anthracene)  &  75\%  & 80\% & 68\%  & 21\%\\
        Peak emission wavelength [nm] & 420    & 425 & 424  & 425        \\ 
        Flash point [$^{\circ}$C]   &54  & 45    & 44 & 115    \\        
        H/C ratio                   & 1.21 & 1.33 & 1.32 & 2.06  \\ 
        Gadolinium content [\%w/w] & -   & -  & 1.5\%   & -   \\
        \hline
    \end{tabular}
    \label{table:ScintillatorProperties}
    \end{table} 

\subsection{Experimental setup}
\label{subsection:ApparatusExperimentalSetup}

\begin{figure}[H]
    \centering
    \includegraphics[width=\textwidth]{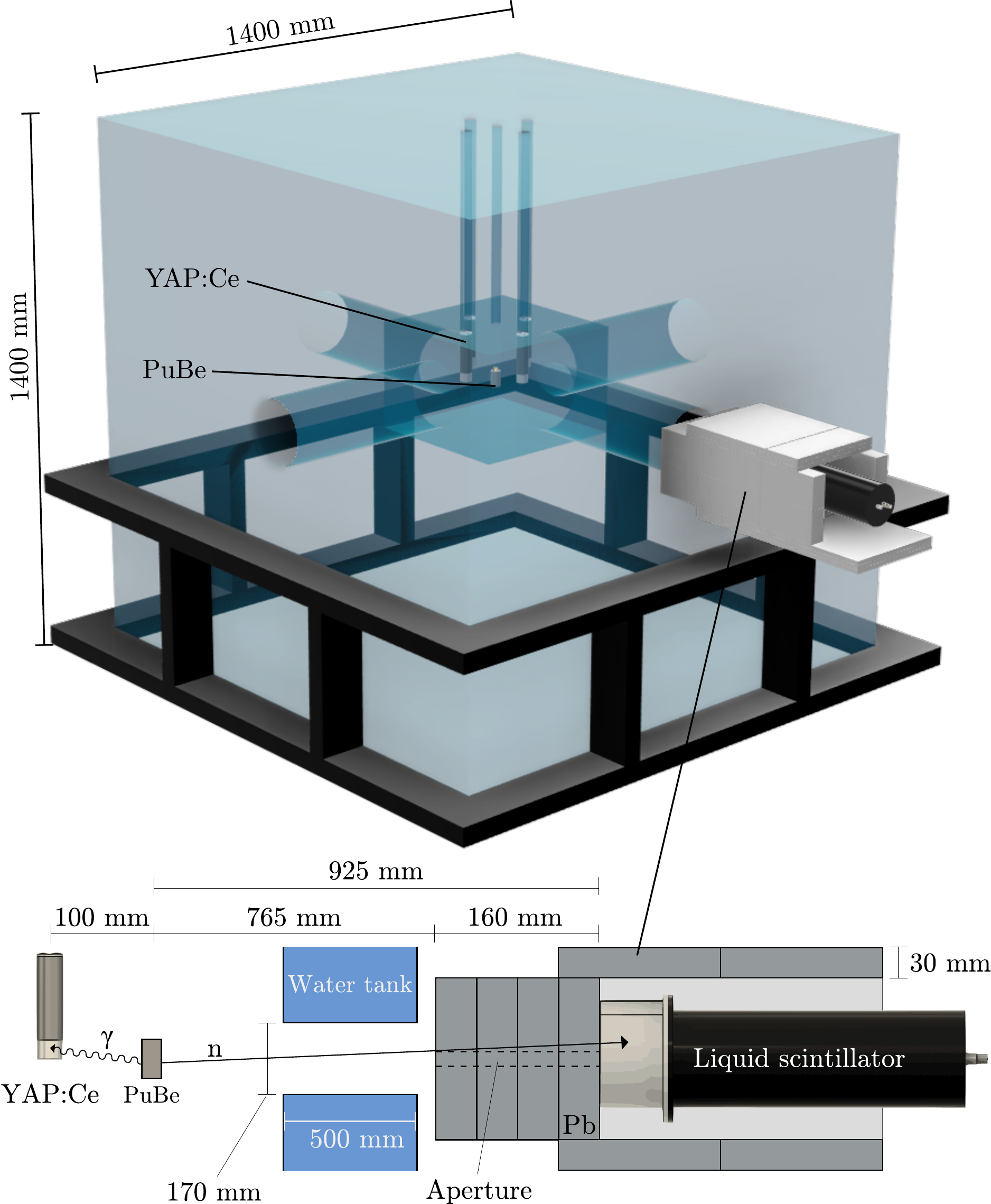}
    \caption{Experimental setup. Top (to scale): 3D rendering of the water tank (Aquarium, blue) and support frame (black) which housed the PuBe source. 
    YAP:Ce detectors and the Pb-shielded liquid scintillator detector are also shown. 
    Bottom (not to scale): Side view of detector setup. 
    The PuBe source emitted correlated 4.44 MeV gamma-ray/fast-neutron pairs. 
    A $\sim$$10\times10$mm$^2$ aperture in the line-of-sight shielding enabled the gamma-flash measurements used to calibrate the TOF measurements.
    For interpretation of the references to color in this figure caption, the reader is referred to the web version of this article.}
    \label{figure:CAD_setup}
\end{figure}

 Figure~\ref{figure:CAD_setup} shows the experimental setup. A water-filled shielding cube known as the ``Aquarium''~\cite{international2020iaea} housed the PuBe source. 
 Each side wall of the cube had a central cylindrical penetration (17\,cm in diameter $\times$ 50\,cm in length) which allowed a mixed beam of fast neutrons and gamma-rays to escape. 
 Four YAP:Ce detectors were placed at a distance of $\sim$10\,cm from the center of the source which was placed at the center of the cube and thus centered on the beam ports. 
 A Pb-shielding hut was constructed outside one of the beam ports. 
 It contained the liquid-scintillator detectors positioned at a distance of 92.5\,cm from the center of the PuBe source. 
 The symmetry axis of the neutron detector was aligned parallel to the beam port and pointed directly at the source. 
 The background rate inside the Pb-shielding hut was measured to be $<$1\,Hz with a 1.5\,in. CeBr$_{3}$ inorganic scintillator detector ($-$600\,V, $-$50\,mV threshold). 
 In comparison, the neutron detectors showed a background rate of $<$100\,Hz ($-$2\,kV, $-$25\,mV threshold). 
 A $\sim$$10\times10$\,mm$^2$ aperture was left in the Pb shielding to allow for the measurement of both line-of-sight low-energy cascade gamma-rays and energetic 4.44\,MeV gamma-rays. 
 
 Two classes of events were of particular interest, see Ref.~\cite{SCHERZINGER201574}: 

\begin{enumerate}
    \item{``tagged-neutron'' events: a fast neutron detected in the neutron detector in correlation with a 4.44\,MeV gamma-ray detected in a YAP:Ce detector.}
    \item{``gamma-flash'' events: a low-energy cascade gamma-ray detected in the neutron detector in correlation with a 4.44\,MeV gamma-ray detected in a YAP:Ce detector.}
\end{enumerate}

\subsection{Electronics and data acquisition}
\label{subection:ApparatusElectronicsDataAcquisition}

\begin{figure}[H]
    \centering
    \includegraphics[width=\textwidth]{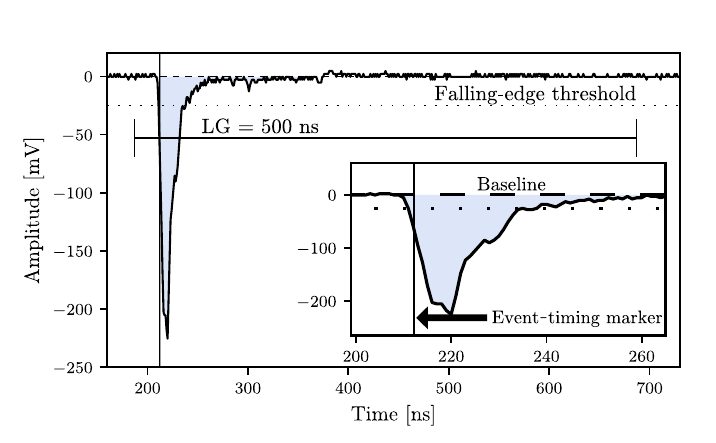}
    \caption{Digitized waveform. The displayed signal has a risetime of  $\sim$5\,ns, a peak amplitude of $\sim$230\,mV and a falltime of  $\sim$50\,ns. The falling-edge trigger set to $-$25\,mV is shown as a dotted line. The event timing marker and the 500\,ns integration window are also shown.}
    \label{figure:waveform}
\end{figure}

Liquid-scintillator pulses were recorded using a CAEN VX1751 Waveform Digitizer~\cite{caen_vx1751}. 
A trigger threshold was set at $-25$\,mV on the falling edge of the pulse.
This started a $1$\,\textmu{}s wide acquisition window over which $10^{3}$ voltage samples were digitized with $10$-bit precision on a dynamic range of $1$\,V. 
Software tools~\cite{nppp2020} for waveform analysis based on the \python~\cite{python} code libraries \texttt{numpy}~\cite{numpy}, \SciPy~\cite{scipy} and \pandas~\cite{pandas} were developed and employed. 
The event-timing marker for each pulse was determined with an interpolating zero-crossover method~\cite{GFKNOLL} which largely removed the time walk associated with the internal falling-edge trigger. 
Figure~\ref{figure:waveform} shows the resulting waveform after the signal baseline was subtracted. 
The effective total signal charge (6.35$\pm$5.5\% fC/channel) was determined by integrating each pulse over 500\,ns starting 25\,ns before the event-timing marker.

\subsection{Scintillation simulation and energy calibration}
\label{subsection:ApparatusScintillationSimulationCalibration}

\subsubsection{Scintillation simulation}
\label{subsubsection:ApparatusScintillationSimulationCalibrationSimulation}

For a particle of energy $E$ that stops in a scintillator, the scintillation light yield is given by
\begin{equation}
L(E) = \int_0^R \frac{dL}{dx} dx,
    \label{eq:total_scintillation_light_yield}
\end{equation}
\noindent
where $\frac{dL}{dx}$ is the scintillation gradient with respect to the path-length increment $dx$ and $R$ is the particle range. 
For minimum-ionizing particles such as the electrons produced by the gamma-ray sources employed here, the scintillation gradient is
\begin{equation}
\frac{dL}{dx} = S \frac{dE}{dx},
    \label{eq:MinimumIonzingScintillationGradient}
\end{equation}
\noindent
where $S$ is the scintillation efficiency and $\frac{dE}{dx}$ is the specific electron energy loss (stopping power). 
For electrons above $\sim$$100$\,keV,  $L(E)$ is linear and it is convenient to express $L$ in terms of $E_{ee}$ (equivalent electron energy, units MeV$_{ee}$). 
In contrast, non minimum-ionizing particles have non-linear scintillation gradients given by the Birks formula~\cite{JBBirks_1951}, which is often modified with the Chou correction~\cite{CHOUPhysRev.87.904} to improve agreement with data at lower energies

\begin{equation}
    \frac{dL}{dx} = S\frac{\frac{dE}{dx}}{1+kB\frac{dE}{dx}+C\left(\frac{dE}{dx}\right)^2}.
    \label{eq:scintillation_light_yield}
\end{equation}

\noindent 
Here, $kB$ is the Birks parameter and $C$ is the Chou correction factor. 
The scintillation light yield is quenched with respect to minimum-ionizing electrons having the same specific energy loss. 

Simulations of the detector response to gamma-rays and neutrons were performed using \geant~\cite{AGOSTINELLI2003250, 1610988}~version 4.10.04 patch~03 (8 February 2019) using a physics list based on the electromagnetic physics classes \texttt{G4EmStandardPhysics} and \texttt{G4EmExtraPhysics}, the hadronic physics class \texttt{FTFP\_BERT\_HP} and optical photon class \texttt{G4OpticalPhysics}. 
Scintillation photons were produced along the tracks of secondary charged particles~\cite{g4bad,g4prm,gumplinger02}, electrons (from gamma-rays) and protons or $^{12}$C (from neutrons). 
Photons which reached the photocathode of the PMT generated photoelectrons with a probability derived from the wavelength-dependent quantum efficiency~\cite{et_9821kb} of the PMT. 
The photoelectron yield as a function of incident energy is effectively a pulse-height distribution which can be compared to the measured data.
Standard \geant~models the scintillation yield without the Chou correction ($C=0$). 
For reproducibility, rather than modifying \geant~to include the $C$ term, an energy-dependence in $kB$ was permitted.

\subsubsection{Energy calibration}
\label{subsubsection:ApparatusScintillationSimulationCalibrationEnergyCalibration}

The light yield produced by gamma-rays in scintillating liquids below pair-production threshold is dominated by Compton scattering due to the low average $Z$ value of the constituent atoms. 
Although pair production becomes increasingly important as the gamma-ray energy increases above threshold, the Compton edge remains a valuable feature for calibration of the pulse-height spectrum. 

The sources listed in Table~\ref{table:GammaRaySources} were placed in front of each neutron detector and spectra were obtained for run times of about one hour per run. 
The measured deadtime and pileup were negligible as the count rates were low ($<1$\,kHz) and gain drift ($\pm$5\%) was corrected for offline. 
Background subtraction was performed after a real-time normalization. 

\begin{table}[H]
    \centering
    \caption{Calibration gamma-ray sources. Distances, gamma-ray energies and Compton-edge energies $E_{\rm CE}$ are listed.}
    \begin{tabular}{r c c c} \hline
                Source     & Distance [cm]& $E_{\gamma}$ [MeV] & $E_{\rm CE}$ [MeV$_{ee}$] \\ \hline
                $^{137}$Cs &  50 &              0.66 &                      0.48 \\
                $^{232}$Th &  50 &              2.62 &                      2.38 \\
$^{241}$Am/$^{9}$Be (AmBe) & 200 &              4.44 &                      4.20 \\ \hline
    \end{tabular}
    \label{table:GammaRaySources}
\end{table}
\noindent
For the full \geant~simulations of the gamma-ray response, the only free parameter was the scale factor necessary to match the distribution of simulated photoelectrons at the photocathode of the PMT to the pulse-height spectrum measured by the detector. 
Smearing due to electronic jitter, extended source and finite detector volume was also included~\cite{MAURITZSON2022165962}.
The simulated locations of the Compton edges were determined by selecting events where the electron energy was less than $2$\,keV from Compton-edge energies.

\begin{figure}[H]
    \centering
    \includegraphics[width=\textwidth]{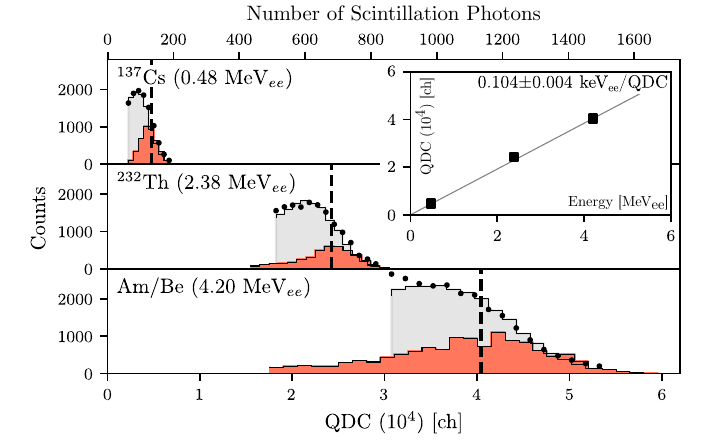}
    \caption{Energy calibration for NE~213A. Measured and simulated Compton distributions for three gamma-ray energies. Main plot: measurement (filled circles), simulation (gray shaded histograms) and simulation with a very restrictive cut on the Compton edge (red shaded histograms). The mean values of the red shaded distributions are shown by the vertical dashed lines. Inset: the resulting QDC calibration. The uncertainties are smaller than the data points. For interpretation of the references to color in this figure caption, the reader is referred to the web version of this article.}
    \label{figure:calibration}
\end{figure}

\subsection{Event selection}
\label{section:ApparatusEvent_selection}
\begin{figure}[H]
    \centering
    \includegraphics[width=\textwidth]{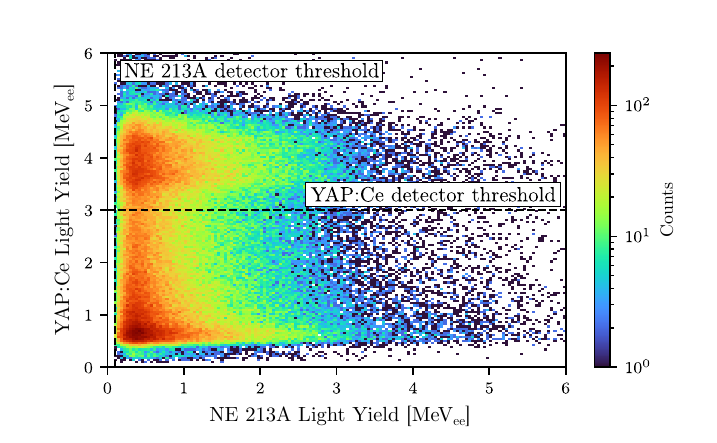}
    \caption{Calibrated scintillation light yields, YAP:Ce and NE~213A. The dashed lines are the detector thresholds, 3\,MeV$_{ee}$ (YAP:Ce) and 100\,keV$_{ee}$ (NE~213A). The events lying above the YAP:Ce threshold are candidate tagged neutrons. Figure from Ref.~\cite{MAURITZSON2022167141}.}
    \label{figure:energy_correlation}
\end{figure}

Figure~\ref{figure:energy_correlation} shows a typical energy-deposition correlation between a YAP:Ce and liquid scintillator (NE~213A) detector. 
The gain of the YAP:Ce detector was set using the full-energy peak of the 1.28\,MeV gamma-ray from $^{22}$Na and the Compton edge of the 4.44\,MeV gamma-ray from PuBe. 
A 3\,MeV$_{ee}$ threshold cut for the YAP:Ce detector allowed for the straightforward selection of 4.44\,MeV gamma-rays, corresponding to neutron emission. 
The intense low-energy gamma-ray field at the center of the water cube prevented selection of lower energy cascade gamma-rays, which in principle could be used to tag higher energy neutrons.
A 100\,keV$_{ee}$ threshold was applied to the NE~213A detector.

\section{Results}
\label{section:Results}

\begin{figure}[H]
    \centering
    \includegraphics[width=\textwidth]{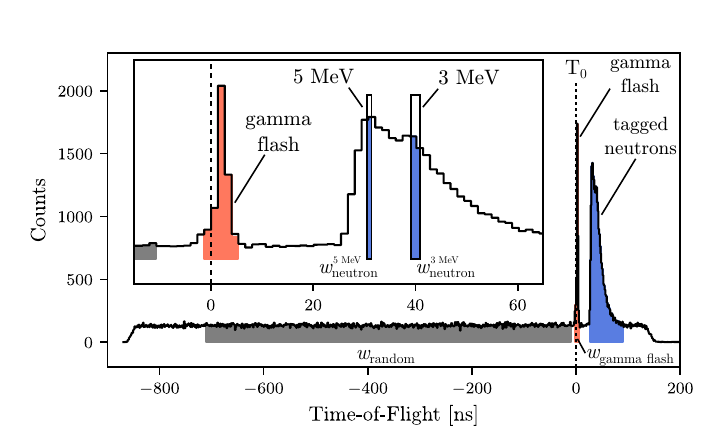}
    \caption{TOF spectrum, NE~213A. Main plot: the full range of the digitization window, displaying $T_0$ (vertical dashed line), gamma flash (sharp red peak) and neutron distributions (broader blue peak). The gray shaded region of the flat background was employed for random subtraction. Inset: region-of-interest. The blue vertical rectangles illustrate the TOF range corresponding to 250\,keV neutron-energy bins centered at 3 and 5\,MeV. For interpretation of the references to color in this figure caption, the reader is referred to the web version of this article.}
    \label{figure:tof_slice}
\end{figure}

Figure~\ref{figure:tof_slice} shows a neutron TOF distribution obtained for a $\sim$96\,cm drift distance between the PuBe source and the NE~213A detector. 
The time $T_0$ located at 0\,ns indicates the instant of emission of the gamma-ray/gamma-ray (gamma flash) or gamma-ray/fast-neutron (tagged neutron) pairs from the PuBe source. 
$T_0$ is extrapolated from the gamma flash timing to the right of $T_{0}$ at $\sim$2.9\,ns. 
The combination of electronic jitter, extended source and finite detector volumes gives rise to the $\sim$1\,ns width of the peak. 
The broad peak starting at $\sim$25\,ns results from tagged neutrons.
The flat distribution corresponds to uncorrelated signals in the YAP:Ce and liquid scintillator. 
The contribution of this random distribution was subtracted from the tagged-neutron distribution using an analysis technique employed for tagged-photon experiments~\cite{OWENS1990574} which considered the ratio of the correlated and random time-window widths. 
Neutron TOF was converted to neutron kinetic energy on an event-by-event basis.

\noindent

\begin{figure}[H]
    \centering
    \includegraphics[width=\textwidth]{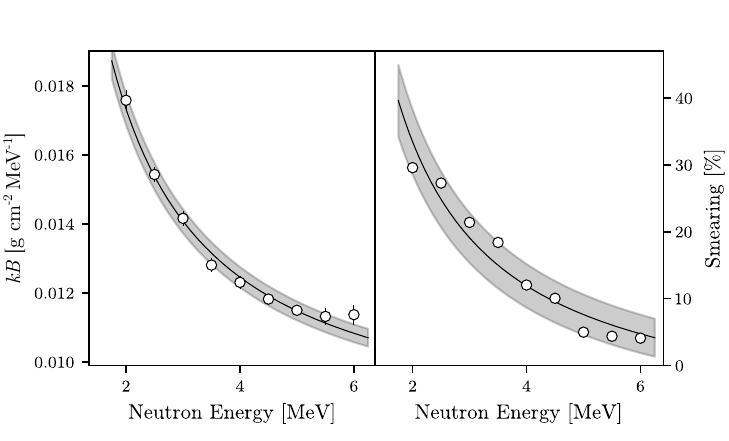}
    \caption{Optimized $kB$ values (left) and smearing values (right), EJ~305. Data points (open circles), fitted trends (solid lines) and uncertainties in the fitted trends (shaded areas) are shown.}
    \label{figure:optimal_kB_and_smearing}
\end{figure}

Standard \geant~does not handle the inclusion of the Chou correction to the Birks formula and thus $kB$ was allowed to vary. 
For each scintillator and each neutron energy bin, the simulation was aligned with the data using a least-squares minimization to obtain the optimum value of $kB$. 
Additional fine-tuning in the agreement was then provided by smearing the simulated scintillation light yield. 
This smearing ranged from $\sim$$35$\% at $2$\,MeV to $\sim$$5$\% at $6$\,MeV for all scintillators. 
The smearing includes effects from non-pointlike source, signal-propagation and electronic noise. 
Figure \ref{figure:optimal_kB_and_smearing} shows the optimal $kB$ and smearing values together with $1/\sqrt{E_n}$ fitted trends. 
Since counting statistics dominates the falloff in the $kB$ and smearing distributions, the $1/\sqrt{E_n}$ dependence shown is anticipated.
The fitted trends are used to generate the scintillation-light yields in the neutron simulations.

\begin{figure}[H]
    \centering
    \includegraphics[width=\textwidth]{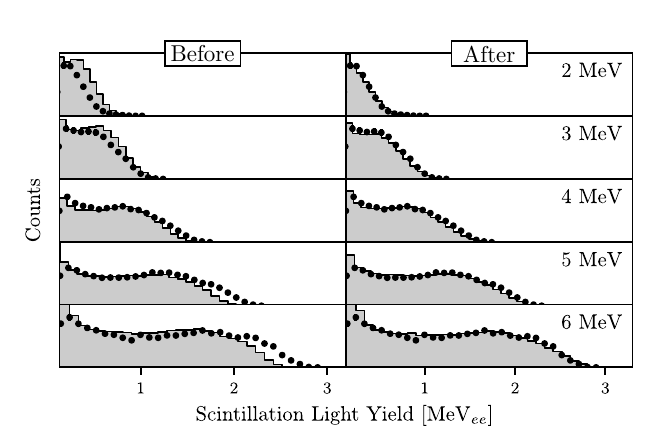}
    \caption{Scintillation light yield, EJ~305. Measured calibrated scintillation light yield (filled circles) and simulations (shaded histograms) are shown before (left) and after (right) $kB$ and smearing optimization. The uncertainties are smaller than the data points.}
    \label{figure:kb_optimization_result}
\end{figure}

Figure~\ref{figure:kb_optimization_result} shows the agreement between data and simulation over the entire energy range before and after $kB$ and smearing optimization. 
Clearly, the energy-dependent $kB$ and smearing optimizations are essential to the reproduction of the data.

\begin{figure}[H]
    \centering
    \includegraphics[width=\textwidth]{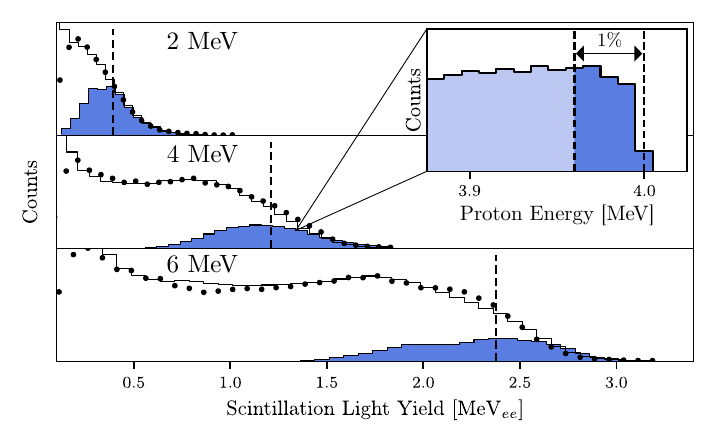}
    \caption{Simulated maximum neutron-energy depositions, EJ~305. Main plot: measured scintillation light yield (filled circles) and full simulations (open histograms) for incident neutron energies 2, 4 and 6\,MeV. 
    Inset: simulated proton recoil energy for a $4$\,MeV pencil neutron beam directed at the center of the detector. The dark shaded region between the vertical dashed lines in the inset corresponds to a $1$\% energy cut and results in the dark shaded simulated scintillation light yield in the middle panel. The SMD locations are shown as vertical dashed lines. The uncertainties are smaller than the data points.}
    \label{figure:neutron_LY_method_1}
\end{figure}

Figure~\ref{figure:neutron_LY_method_1} shows the neutron scintillation light yield from EJ~305 for the measured data, the full simulation and the simulated maximum neutron-energy deposition (SMD). 
To determine the SMD, a point source, non-divergent, monoenergetic (pencil) neutron beam was directed at the center of the scintillator cell. 
For each incident neutron-beam energy, the energy deposited by recoiling protons as the neutrons traversed the cell was recorded. 
A $1$\% cut on the high-energy edge of the proton-energy distribution was then enforced to populate the scintillation light-yield spectra corresponding to the SMD. 
To exclude tail contributions, the SMD distributions were then fitted with a Gaussian function and values within $\pm 3\sigma$ of the mean were used to determine the average peak position.

\begin{figure}[H]
    \centering
    \includegraphics[width=\textwidth]{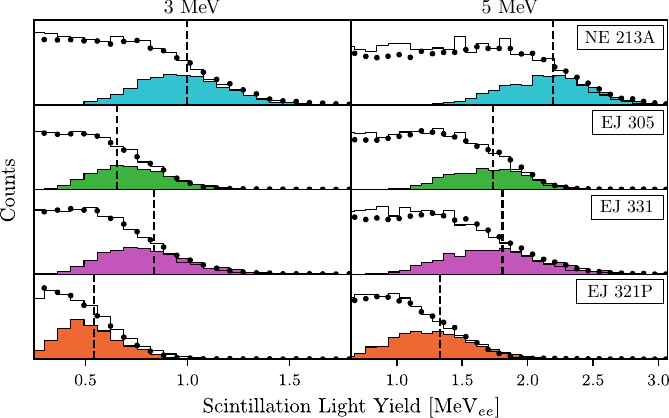}
    \caption{Simulated maximum depositions, all scintillators. Measured scintillation light yields (filled circles) and full simulations (open histograms) are shown together with the SMD simulations (colored histograms). The SMD locations are shown as vertical dashed lines. The uncertainties are smaller than the data points.}
    \label{figure:LY_all_detectors}
\end{figure}

Figure~\ref{figure:LY_all_detectors} shows a comparison between the data, the full simulations, the SMD simulations and the corresponding SMD values for all scintillators for 250\,keV neutron-energy bins centered at 3 and 5\,MeV.
The agreement between the simulation and the data was excellent.

Phenomenological parameterizations of neutron scintillation-light yield based upon the specific energy loss of protons ($E_p$) have been used to characterize measured neutron spectra.
The correlation between recoiling electron ($L_{ee}$) and quenched recoiling proton ($L(E_p)$) scintillation light yields was determined by Cecil~\etal~\cite{CECIL1979439} for NE~213 to be

\begin{equation}
    L(E_p) = L_{ee} = K\left[p_1E_p-p_2\left(1-e^{-p_3E_p^{p_4}}\right)\right],
    \label{eq:cecil}
\end{equation}
while Kornilov~\etal~\cite{KORNILOV2009226} suggested
\begin{equation}
    L(E_p) = L_{ee} = L_0 \frac{E^2_p}{E_p+L_1}.
    \label{eq:kornilov}
\end{equation}
In Eqs. \ref{eq:cecil} and \ref{eq:kornilov}, $K$ and $L_0$ are adjustable scaling parameters and $p_{1-4}$ and $L_1$ are material-specific light-yield parameters. 
The maximum energy the incident neutron can transfer to the recoiling proton in a single scatter may be determined using three different methods to locate the high-energy edge of the scintillation distribution (see for example Ref.~\cite{JuliusScherzinger2016}):

\begin{enumerate}
    \item The half-height (HH) method~\cite{NAQVI1994156} involves fitting a Gaussian function to the edge of the recoil-proton distribution and selecting the half maximum as the location of the maximum proton-energy transfer.
    \item The turning-point (TP) method also involves fitting a Gaussian function, but here the minimum of the first derivative of the function is selected as the maximum proton-energy transfer.
    \item The first-derivative (FD) method~\cite{KORNILOV2009226} involves taking the first derivative of the distribution and selecting the minimum point as the maximum proton-energy transfer. 
    In this work, the first derivative was evaluated by considering 5 adjacent bins above and below each data point (11 bins total).
\end{enumerate}

\begin{figure}[H]
    \centering
    \includegraphics[width=\textwidth]{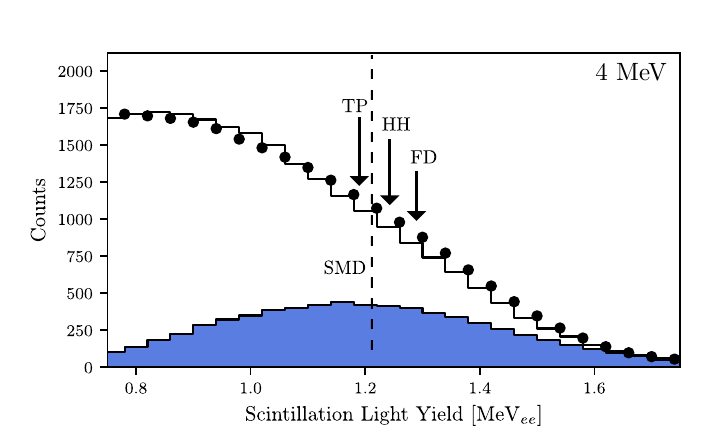}
    \caption{Simulated maximum depositions and proton edge locations, EJ~305. Measured scintillation light yield (filled circles) and full simulation (open histogram) together with the SMD simulation (shaded histogram) and the SMD location (dashed line) are shown. The vertical arrows indicate the maximum recoil proton edges as predicted by the HH, TP and FD methods.}
    \label{figure:neutron_LY_methods}
\end{figure}

For the purposes of comparison, the SMD employed in the simulation-driven analyses of scintillation light yield may be compared with the maximum proton recoil edge employed in the HH, TP and FD methods.
Figure~\ref{figure:neutron_LY_methods} shows the scintillation light yields from $4$\,MeV neutrons with the SMD and HH, TP and FD recoil-proton edge locations indicated. 

\noindent While the HH, TP and FD locations generally have the same relative locations with respect to one another regardless of the neutron energy bin, the relative location of the SMD varies with neutron energy.

For NE~213A and EJ~305, parameterization coefficients corresponding to NE~213~\cite{CECIL1979439} (NE~213A equivalent) and EJ~309~\cite{ENQVIST201379} (EJ~305 equivalent) were employed to determine the light-yield curves corresponding to Eq.~\ref{eq:cecil} (Cecil~\etal). 
The base organic in EJ~331 was assumed to be EJ~309, see Table~\ref{table:fixed_parameters_organics}. 
The parameterization coefficients $p_{1-4}$ for NE~213A, EJ~305, EJ~331 and EJ~321P were also determined by fitting to the maximum recoil proton-edge distributions for the HH, TP and FD methods.
The fitted results for $p_{1-4}$ from the HH, TP and FD distributions were averaged and fixed as constants. 
$K$ was subsequently determined with these constants.  

The $L_1$ coefficients for all four scintillators were similarly determined by fitting to the data using Eq.~\ref{eq:kornilov} (Kornilov~\etal).
First, the HH, TP and FD neutron scintillation light yields were fitted allowing both $L_0$ and $L_1$ to vary. 
The resulting $L_1$ parameters for HH, TP and FD were then averaged and fixed as an $L_1$ constant. 
In comparison, Scherzinger~\etal~\cite{JuliusScherzinger2016} report $L_1$~$=$~2.48 for NE~213 and Enqvist~\etal~\cite{ENQVIST201379} report $L_1$~$=$~5.95 for EJ~309. 

\begin{table}[H]
\centering
\footnotesize
    \caption{Scintillation parametrization coefficients. Fitted coefficients are shown with errors while fixed parameters are given in parenthesis.}
    \begin{tabular}{lcccc|c}
   \multicolumn{5}{c}{Eq.~\ref{eq:cecil}, Cecil~\etal} & \multicolumn{1}{c}{Eq.~\ref{eq:kornilov}, Kornilov~\etal} \\\hline
    Scintillator & $p_1$ & $p_2$ & $p_3$ & $p_4$ & $L_1$ \\\hline
   NE~213A (NE~213) & 0.65 $\pm$ 0.02 (0.83) & 0.96 $\pm$ 0.12 (2.82) &  0.42 $\pm$ 0.08 (0.25) &  1.72 $\pm$ 0.36 (0.93) &  3.67 $\pm$ 0.19\\
   EJ~305 (EJ~309) & 0.56 $\pm$ 0.01 (0.817) & 0.99 $\pm$ 0.08 (2.63) &  0.44 $\pm$ 0.05 (0.297) &  1.55 $\pm$ 0.24 (1) &  6.55 $\pm$ 0.38 \\
   EJ~331 (EJ~309) & 0.58 $\pm$ 0.01 (0.817) & 1.06 $\pm$ 0.08 (2.63) &  0.29 $\pm$ 0.03 (0.297) &  1.83 $\pm$ 0.20 (1) & 5.34 $\pm$ 0.48 \\
   EJ~321P &  0.43 $\pm$ 0.01 &  0.77 $\pm$ 0.07 &  0.26 $\pm$ 0.07 &  2.14 $\pm$ 0.43 & 6.68 $\pm$ 0.82 \\\hline
    \end{tabular}
    \label{table:fixed_parameters_organics}
\end{table}

\begin{figure}[H]
    \centering
    \includegraphics[width=\textwidth]{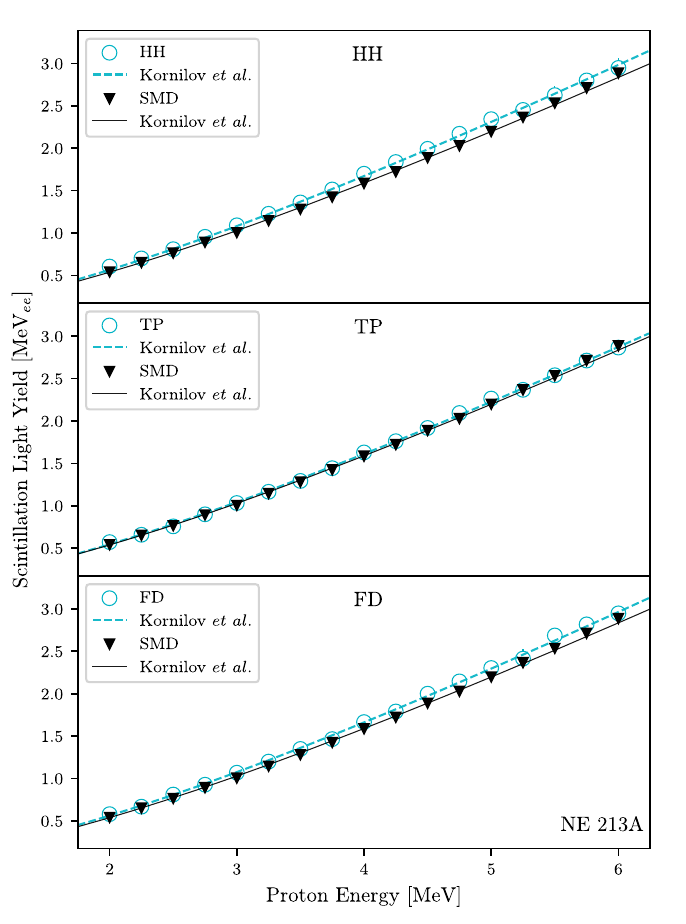}
\caption{Comparison of recoil-proton light yield, NE~213A. Results have been obtained using the SMD (filled triangles, identical in all panels), HH (top panel, open circles), TP (middle panel, open circles) and FD (bottom panel, open circles) methods. The Kornilov~\etal~parameterizations are shown for the HH, TP and FD methods (dashed lines) while the Kornilov fit for the SMD method is the solid lines, again identical in all panels. The uncertainties are smaller than the data points.}
\label{figure:NE213A_result}
\end{figure}

Figure~\ref{figure:NE213A_result} shows light yield as a function of recoil proton energy for NE~213A. 
The SMD method for determining the maximum recoil proton edge is compared with the HH, TP and FD methods.
A summary of the fixed parameters employed in the fitted functions may be found in Table~\ref{table:fixed_parameters_organics}.
The scintillation light yield increases as a function of recoil proton energy, but not linearly due to quenching.  
The TP approach reproduces the SMD results well. 
There is little sensitivity when the HH and FD methods are used to determine the recoil proton edge, and both overestimate the light yields by up to $\sim$6\%.

\begin{figure}[H]
    \centering
    \includegraphics[width=\textwidth]{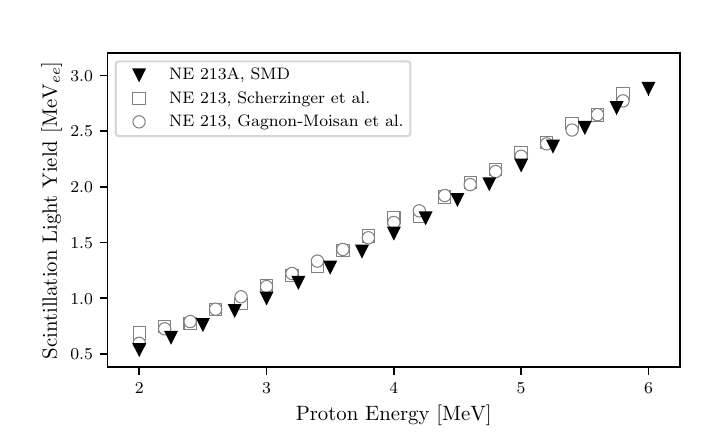}
    \caption{Comparison of recoil-proton light yield measurements, NE~213A. Results include SMD approach (filled triangles), Gagnon-Moisan~\etal~\cite{gagnon2012results} (open circles) and Scherzinger~\etal~\cite{JuliusScherzinger2016} (open squares). The uncertainties are smaller than the data points.}
    \label{figure:NE213A_comparison}
\end{figure}

Figure~\ref{figure:NE213A_comparison} presents a comparison between the NE~213A SMD results detailed above and the scintillation light yield for NE~213 measured by both Gagnon-Moisan~\etal~\cite{gagnon2012results} and Scherzinger~\etal~\cite{JuliusScherzinger2016}. 
Agreement between the data sets and the SMD prescription is very good. The classic scintillator NE~213A appears to be well understood in this energy region.

\begin{figure}[H]
    \centering
    \includegraphics[width=\textwidth]{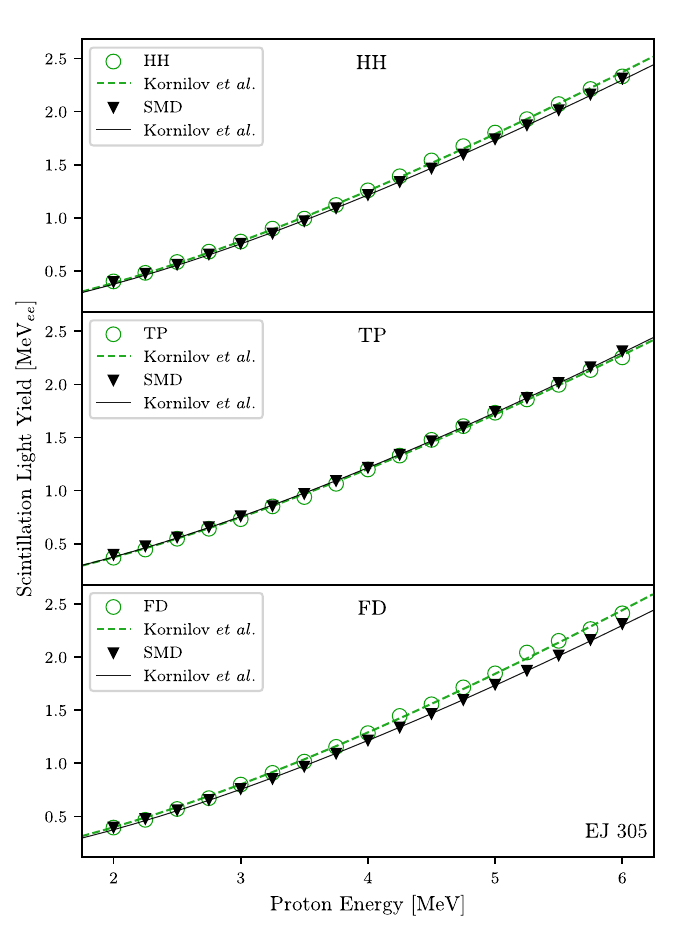}
\caption{Comparison of recoil-proton light yield, EJ~305. Results have been obtained using the SMD (filled triangles, identical in all panels), HH (top panel, open circles), TP (middle panel, open circles) and FD (bottom panel, open circles) methods. The Kornilov~\etal~parameterizations are shown for the HH, TP and FD methods (dashed lines) while the Kornilov fit for the SMD method is the solid lines, again identical in all panels. The uncertainties are smaller than the data points.}
    \label{figure:EJ305_result}
\end{figure}

Figure~\ref{figure:EJ305_result} shows light yield for EJ~305 as a function of recoil proton energy for the SMD and HH, TP and FD methods. 
Again, the TP approach reproduces the SMD results well. 
There is little sensitivity when the HH and FD methods are used to determine the recoil proton edge, and both overestimate the light yields by up to $\sim$8\%.

\begin{figure}[H]
    \centering
    \includegraphics[width=\textwidth]{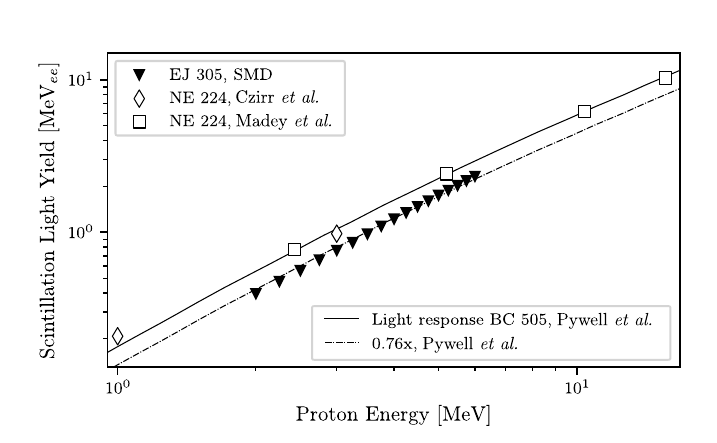}
    \caption{Calibrated neutron scintillation light-yield comparison, EJ~305 (this work), BC~505 and NE~224. The NE~224 results of Czirr~\etal~(open diamonds) and Madey~\etal~(open squares) are shown together with the EJ~305 SMD prescription (filled triangles). The uncertainties are smaller than the data points.}
    \label{figure:EJ305_comparison}
\end{figure}

Figure~\ref{figure:EJ305_comparison} presents a comparison between the EJ~305 SMD prescription and the scintillation light yields for NE~224 (EJ~305 equivalent) measured by both Czirr~\etal~\cite{CZIRR1964226} and Madey~\etal~\cite{MADEY1978445} together with the parameterization for BC~505 (EJ~305 equivalent) determined by Pywell~\etal~\cite{pywell2006light}. 
The dash-dotted line represents the Pywell~\etal~parameterization scaled by 0.76, determined by least-squares minimization.
The scaled parameterization underestimates the scintillation light yields measured with NE~224 and shows a slightly weaker scintillation light-yield gradient than the SMD prediction. 
The comparison between NE~224, BC~505 and EJ~305 may not be optimal but nevertheless provides insight into the behavior of these closely related organics.

\begin{figure}[H]
    \centering
    \includegraphics[width=\textwidth]{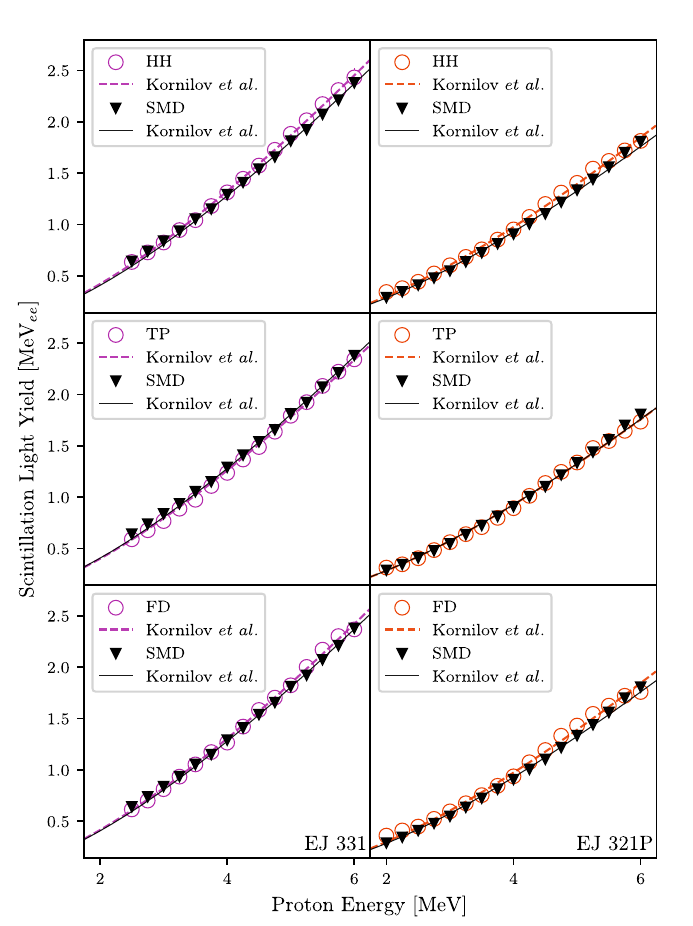}
    \caption{Comparison of recoil-proton light yield, EJ~331 and EJ~321P. Results have been obtained using the SMD (filled triangles, identical in all panels), HH (top panel, open circles), TP (middle panel, open circles) and FD (bottom panel, open circles) methods. The Kornilov~\etal~parameterizations are shown for the HH, TP and FD methods (dashed lines) while the Kornilov fit for the SMD method is the solid lines, again identical in all panels. The uncertainties are smaller than the data points.}
    \label{figure:EJ331_EJ321P_result}
\end{figure}

Figure~\ref{figure:EJ331_EJ321P_result} shows light yield as a function of recoil proton energy for EJ~331 and EJ~321P.
The manner of presentation is identical to that employed for Figs.~\ref{figure:NE213A_result} and~\ref{figure:EJ305_result} and the trends in the results are similar. 
The TP method does an excellent job of reproducing the SMD results for both scintillators while the HH and FD methods overestimates the light yields by up to $\sim$5\% (EJ~331) and $\sim$7\% (EJ~321P), respectively.

Table~\ref{tab:fitted_parameters_organics} presents a summary of the $K$ and $L_0$ results extracted from fitting the Cecil~\etal~and Kornilov~\etal~curves (using the fixed parameters described in Table~\ref{table:fixed_parameters_organics}) to the scintillation light-yield data and SMD results shown in Figs.~\ref{figure:NE213A_result}, \ref{figure:EJ305_result} and \ref{figure:EJ331_EJ321P_result}.
While generally not consistent within uncertainty, there is little to distinguish between the $K$ and $L_0$ coefficients resulting from the different methods for determining the recoil proton edges. 
The NE~213A results are systematically $\sim$3\% lower for $K$ and about $\sim$8\% higher for $L_0$ than those measured for NE~213 by Scherzinger~\etal~\cite{JuliusScherzinger2016}. 
This is due to the different value of $L_1$ being employed in this work. 
The $K$ and $L_0$ coefficients corresponding to the SMD result are systematically lower than the HH, TP and FD results by $\sim$5\%. 

\begin{table}[H]
    \centering
    \small
    \caption{Fitted scintillation parameterization coefficients. The fixed parameters employed in the fits may be found in Table~\ref{table:fixed_parameters_organics}.}
    \begin{tabular}{ l c | c r | c r | c r }\hline
    Scintillator & Edge & $K_\text{fitted}$ & $\chi^2$/d.o.f. & $K_\text{fixed}$ & $\chi^2$/d.o.f. & $L_0$ & $\chi^2$/d.o.f.  \\\hline
    NE213A  & HH &0.99 $\pm$ 0.01 &0.2& 1.02 $\pm$ 0.01 & 1.6 & 0.80 $\pm$ 0.01 & 2.4\\
            & TP &0.95 $\pm$ 0.01 &2.1& 0.98 $\pm$ 0.01 & 4.7 & 0.77 $\pm$ 0.01 & 1.8\\
            & FD &0.98 $\pm$ 0.01 &1.3& 1.01 $\pm$ 0.01 & 1.8 & 0.80 $\pm$ 0.01 & 0.7\\
           & SMD &0.94 $\pm$ 0.01 &1.5& 0.97 $\pm$ 0.01 & 2.2 & 0.76 $\pm$ 0.01 & 0.4\\\hline
    EJ305   & HH &1.00$\pm$ 0.01 &2.7& 0.87 $\pm$ 0.01 & 0.4 & 0.83 $\pm$ 0.01 & 0.7\\
            & TP &0.96$\pm$ 0.01 &2.9& 0.84 $\pm$ 0.01 & 1.3 & 0.79 $\pm$ 0.01 & 0.7\\
            & FD &1.03$\pm$ 0.01 &1.7& 0.90 $\pm$ 0.01 & 1.1 & 0.85 $\pm$ 0.01 & 0.7\\
           & SMD &0.97$\pm$ 0.01 &1.2& 0.84 $\pm$ 0.01 & 0.1 & 0.80 $\pm$ 0.01 & 0.3\\\hline
    EJ331   & HH &1.05$\pm$ 0.01 &6.1& 0.91 $\pm$ 0.01 & 1.1 & 0.77 $\pm$ 0.01 & 0.6\\
            & TP &1.00$\pm$ 0.01 &1.6& 0.87 $\pm$ 0.01 & 0.7 & 0.74 $\pm$ 0.01 & 1.4\\
            & FD &1.04$\pm$ 0.01 &1.5& 0.90 $\pm$ 0.01 & 3.5 & 0.76 $\pm$ 0.01 & 2.9\\
           & SMD &1.02$\pm$ 0.01 &4.2& 0.88 $\pm$ 0.01 & 4.7 & 0.75 $\pm$ 0.01 & 3.2\\\hline
    EJ321P  & HH &0.99$\pm$ 0.01 &2.0& 1.02 $\pm$ 0.01 & 1.3 & 0.65 $\pm$ 0.01 & 7.2\\
            & TP &0.94$\pm$ 0.01 &3.0& 0.97 $\pm$ 0.01 & 1.2 & 0.62 $\pm$ 0.01 & 3.2\\
            & FD &0.99$\pm$ 0.01 &1.0& 1.02 $\pm$ 0.01 & 1.7 & 0.65 $\pm$ 0.01 & 4.5\\
           & SMD &0.95$\pm$ 0.01 &6.7& 0.97 $\pm$ 0.01 & 3.2 & 0.62 $\pm$ 0.01 & 1.8\\\hline
\end{tabular}
    \label{tab:fitted_parameters_organics}
\end{table}

\section{Summary and discussion}
\label{section:Summary_and_discussion}

Beams of energy-tagged neutrons from 2--6\,MeV provided by a PuBe source have been used to perform a systematic study of the scintillation light-yield response of the scintillators NE~213A, EJ~305, EJ~331 and EJ~321P. 
Neutron tagging exploits the $\alpha$~$+$~$^{9}$Be~$\rightarrow$~$^{12}$C~$+$~$n$~$+$~$\gamma$(4.44\,MeV) reaction, with the gamma-rays providing a reference for measuring the TOF of the correlated neutron.
The PuBe source and YAP:Ce gamma-ray detectors were placed within a water-filled shielding cube. 
The cube employed cylindrical ports to define beams of gamma-rays and fast neutrons. 
Pb shielding attenuated the majority of the direct gamma-rays from the PuBe and the background gamma-rays from the room (Fig.~\ref{figure:CAD_setup}). 
The analog signals from the detectors were digitized on an event-by-event basis, with the event-timing marker determined using an interpolating zero-crossover method (Fig.~\ref{figure:waveform}). 
Energy calibration of the resulting spectra was performed using a 
\geant~model of the liquid scintillator to locate the position of the Compton edge in the measured gamma-ray spectra from gamma-ray sources (Fig.~\ref{figure:calibration}). 
The correlation between the energy registered in a YAP:Ce gamma-ray detector and the energy deposited in a liquid-scintillator was used to select tagged events (Fig.~\ref{figure:energy_correlation}). 
Neutron energies were determined using the TOF method and the data were corrected for random background (Fig.~\ref{figure:tof_slice}). 

Neutron scintillation-light yield was simulated using the same \geant~model and matched to the data by allowing an energy dependence in the Birks parameter (Fig.~\ref{figure:kb_optimization_result}). 
The simulated yield corresponding to the maximum neutron-energy deposition was determined with a very strict cut on the deposited neutron energy (Fig.~\ref{figure:neutron_LY_method_1}). 
The method worked very well (Fig.~\ref{figure:LY_all_detectors}). 
The relationship between the simulated maximum deposition (SMD) light yield and scintillation light yields corresponding to the maximum proton recoil edge for the HH, TP and FD methods was determined (Fig.~\ref{figure:neutron_LY_methods}). 
Data and simulation for NE~213A agreed very well (Fig.~\ref{figure:NE213A_result}) and nicely reproduced existing results (Fig.~\ref{figure:NE213A_comparison}). 
Data and simulation for EJ~305 also agreed well (Fig.~\ref{figure:EJ305_result}), however they showed a steeper energy dependence compared with the parameterization of existing data (Fig.~\ref{figure:EJ305_comparison}). 
Results were obtained for EJ~331 and EJ~321P (Fig.~\ref{figure:EJ331_EJ321P_result}) for the first time to the knowledge of the authors. 

The neutron-tagging technique facilitates the measurement of energy-dependent scintillator response using radioactive neutron sources.
An accelerator-based neutron generator (see Refs.~\cite{frost2022development,frost2023compact}) could be used to extend the results to higher neutron energies, as the tagged neutron energy range provided by the PuBe
source is relatively small.
The \geant~simulation developed and tested here provides valuable insight into the scintillation light production mechanism and the propagation of the scintillation light within the detector assembly. 
This allows for a precise determination of the scintillation-light yield for each of the scintillators.

\section*{Acknowledgements}
\label{Section:Acknowledgements}
Support for this project was provided by the European Union via the Horizon 2020 BrightnESS Project (Proposal ID 676548) and the UK Science and Technology Facilities Council (Grant No. ST/P004458/1).
\newpage

\bibliographystyle{elsarticle-num}

\end{document}